\documentclass[aps,prb,twocolumn,showpacs,groupedaddress]{revtex4}
\usepackage{graphics,bm}
\usepackage{amssymb}
\usepackage{amsmath}
\usepackage{epsfig}
\usepackage{epsf}
\usepackage{verbatim}

\begin{document}

\title{Current-driven and field-driven domain walls at nonzero temperature}

\author{M.E. Lucassen}
\email{m.e.lucassen@uu.nl}

\author{H.J. van Driel}

\author{C. Morais Smith}

\author{R.A. Duine}

\affiliation{Institute for Theoretical Physics, Utrecht
University, Leuvenlaan 4, 3584 CE Utrecht, The Netherlands}
\date{\today}

\begin{abstract}
We present a model for the dynamics of current- and field-driven
domain-wall lines at nonzero temperature. We compute
thermally-averaged drift velocities from the Fokker-Planck
equation that describes the nonzero-temperature dynamics of the
domain wall. As special limits of this general description, we
describe rigid domain walls as well as vortex domain walls. In
these limits, we determine also depinning times of the domain wall
from an extrinsic pinning potential. We compare our theory with
previous theoretical and experimental work.
\end{abstract}


\maketitle

\def\G{{\rm G}}
\def\L{{\rm L}}
\def\R{{\rm R}}
\def\s{{\rm s}}
\def\p{{(+)}}
\def\m{{(-)}}
\def\K{{\rm K}}
\def\F{{\rm F}}
\def\kB{k_{\rm B}}
\def\ii{{\rm i}}
\def\d{d}
\def\om{\mathbf{\Omega}}
\def\x{{(\vec{x},t)}}

\section{Introduction}\label{section:introduction}

Current-driven domain wall motion was first predicted and observed
by Berger in the eighties.\cite{berger1984, berger1985} It was not
until the discovery, in the nineties, of the spin transfer torque
mechanism,\cite{berger1996, slonczewski1996} that research on
current-driven domain walls took off. Spin transfer torques on a
domain wall can be understood on an intuitive level: the electrons
which constitute the current have spin, and this spin rotates when
it passes through the domain wall, as it aligns with the domain
wall magnetization. By conservation of spin, there is an opposite
torque on the magnetization of the domain wall, which leads to a
net displacement of the domain wall in the same direction as the
electric current. Later, to explain some discrepancies with
experiments, a so-called dissipative spin transfer torque
(sometimes referred to as the non-adiabatic spin transfer torque)
was added to the model.\cite{zhang2004, barnes2005} The value of
the dimensionless parameter $\beta$, which gives the strength of
this torque, has been the subject of much debate. By now it is
generally accepted that $\beta$ is of the same order as $\alpha$,
the Gilbert damping parameter, but not necessarily equal to
it.\cite{tserkovnyak2006,kohno2006,piechon2007,duine2007a,
heyne2008} Furthermore, neither $\beta$ nor $\alpha$ needs to be
constant. They depend on the properties of the material and are
most likely also temperature dependent.

Several properties of current-driven domain walls have been
studied. One particular subject of interest is the velocity of the
domain wall. The effect of an external pinning potential and the
depinning behavior, both without and with thermal fluctuations,
were investigated experimentally.\cite{ravelosona2005, petit2007,
bruno1999} There is also interest in more complex, higher
dimensional domain wall models, like vortex
walls.\cite{kruger2007, he2006, shibata2006} These are especially
attractive from an experimental point of view since their
dynamics, such as precession in a potential and transformations
between vortex walls and transverse walls, can be directly
observed.\cite{klaui2005, hayashi2006, heyne2008, bolte2008}

With a few
exceptions,\cite{duine2008,duine2007b,martinez2007,tatara2005,lecomte2009}
most theoretical papers are restricted to the zero-temperature
case. This is unfortunate, since experiments on current-driven
domain walls are usually done at room temperature, and the
relatively large current will heat up the sample even further.
Furthermore, especially in the presence of a pinning potential,
thermal fluctuations are anything but negligible. Domain pinning
can be necessary, for instance to precisely locate a domain wall,
but thermal depinning can also be useful to lower the critical
current. This means that it is very important to understand the
influence of thermal fluctuations on the behavior of the domain
wall precisely and thoroughly. Here, we present a unified picture
of previous work on domain wall motion at nonzero temperature that
involved two of us, as well as new results following from this
unified picture.

We start this paper with the Landau-Lifschitz-Gilbert equation,
including both the reactive and the dissipative spin transfer
torques. In Section \ref{section:domain-wall lines},  we apply a
variational principle to derive the general equations of motion
for a domain-wall line including thermal fluctuations. In Section
\ref{section:rigid domain walls} we investigate the velocity of
current-driven and field-driven rigid domain walls at nonzero
temperature, while in Section \ref{section:vortex domain walls} we
look at vortex domain walls in more detail. Both the rigid and the
vortex domain wall are special cases of the general description
given in Section \ref{section:domain-wall lines}. In both
sections, we start with the zero-temperature case, and then
investigate the influence of thermal fluctuations. Pinning
potentials are included in both models, and we examine thermal
depinning. Each section is divided into three subsections: one in
which the model is described, one in which we present our results,
and a final one to make a comparison with other work.

\section{Domain-wall lines}\label{section:domain-wall lines}

In this section, we derive the equations of motion for a
domain-wall line in a more detailed way than in previous
work\cite{duine2008} and present new results obtained from this
model. We derive the Fokker-Planck equation of the system, and
determine the stochastic behavior under the influence of
temperature. The model is then considered in the absence of
extrinsic pinning so that there is only intrinsic pinning due to
magnetic anisotropy. We determine the drift velocity of the domain
wall as a function of the current through the system. Finally, we
compare our model with other theoretical and experimental work
available in the literature.

\subsection{Model}

Magnetization dynamics including spin-transfer torques are
described by the Landau-Lifschitz-Gilbert equation
\cite{slonczewski1996,berger1996}
\begin{align}\label{eq:landaulifschitzgilbert}
\left(\frac{\partial}{\partial
t}+\vec{v}_\s\!\cdot\!\vec\nabla\right)\om=\om\times\mathbf{H}
-\alpha\om\times\left(\frac{\partial}{\partial t}
+\frac{\beta}{\alpha}\vec{v}_\s\!\cdot\!\vec\nabla\right)\om\;,
\end{align}
where $\mathbf{\Omega}$ is a unit vector in the direction of the
magnetization.  In this expression, the first term on the r.h.s.
contains a contribution due to the effective field, which is
written as the functional derivative of the micromagnetic energy
functional of the system and an external magnetic field
$\mathbf{H}\x=-\delta E_{\rm
MM}[\om\x]/\hbar\delta\om\x+g\mathbf{B}_{\rm ext}\x/\hbar$. For
clarity, we denote positions in three-dimensional space with an
arrow, and the directions of the magnetization by bold symbols.
The second term describes Gilbert damping, which is characterized
by the dimensionless parameter $\alpha$. The term proportional to
$\vec{v}_\s$ on the l.h.s. is the reactive spin-transfer torque.
The term proportional to $\beta \vec{v}_\s$ on the r.h.s.
corresponds to the dissipative spin-transfer torque and is
characterized by the dimensionless parameter $\beta$. The velocity
$\vec{v}_\s$ is given by $\vec{v}_\s=-a^3P\vec{J}_{\rm c}/|e|$,
where $a$ is the lattice constant, $\vec{J}_{\rm c}$ is the charge
current and $P$ is its spin polarization. To facilitate a
variational approach, we note that the Landau-Lifschitz-Gilbert
equation is obtained from
\begin{align}\label{eq:eulerlagrange}
\frac{\delta S}{\delta\phantom{\lefteqn{\dot\om}}\om\x}
=\frac{\delta R}{\delta \dot{\om}\x}\;,
\end{align}
where $R$ is a dissipation functional, and where $S=S_0+S_{\rm
drive}$ denotes the action of the system. The dot is a time
derivative $\dot\om\equiv\partial\om/\partial t$. The action for
the magnetization dynamics in the absence of current and field is
given by
\begin{align}\label{eq:actionint}
S_0&[\om\x]=\int \!d t\!\int\frac{d^3x}{\;a^3}
\bigg\{\hbar\mathbf{A}(\om\x)\cdot\frac{\partial\om\x}{\partial t
}\nonumber\\
&+J\om\x\cdot\vec{\nabla}^2 \om\x
-K_\perp\Omega_y^2\x+K_z\Omega_z^2\x\bigg\}\nonumber\\
\equiv& \int\!d t\bigg\{\!\int\frac{d^3x}{\;a^3}
\hbar\mathbf{A}(\om\x)\cdot\frac{\partial\om\x}{\partial t}
-E_{\rm MM}[\om\x]\bigg\}.
\end{align}
In this expression, $J$ is the spin stiffness, $K_\perp>0$ and
$K_z>0$ are the hard- and easy-axis anisotropy, respectively. The
function $\mathbf{A}(\om)$ is the vector potential of a magnetic
monopole which obeys
$\boldsymbol\nabla_\om\times\mathbf{A}(\om)=\om$ and is required
to reproduce the precessional motion of $\om\x$ around the
effective field.

The external field $\mathbf{B}_{\rm ext}\x$ and the reactive
spin-transfer torque are determined from the action $S_{\rm
drive}[\om\x]$. We take the external magnetic field to be
spatially homogeneous and time independent $\mathbf{B}_{\rm
ext}\x=\mathbf{B}_{\rm ext}$. The action is given by
\begin{align}\label{eq:actionext}
S_{\rm drive}[\om\x]= \int d t\int&\frac{d^3x}{a^3}\bigg\{
g\mathbf{B}_{\rm
ext}\cdot\om\x\nonumber\\
+&\mathbf{A}(\om)\cdot\left(\vec{v}_\s\cdot\vec\nabla\right)\om\x\bigg\}\;,
\end{align}
with $g$ a positive constant. The dissipation functional that
describes the dissipative spin-transfer torque and the Gilbert
damping is written as
\begin{align}\label{eq:dissipation}
R[\om\x]=\frac{\alpha\hbar}{2}\int d t\int\frac{d^3
x}{a^3}\left[\left(\frac{\partial}{\partial t}
+\frac{\beta}{\alpha}\vec{v}_\s\!\cdot\!\vec\nabla\right)\om\x\right]^2\!\!.
\end{align}

As the last ingredient, we take into account thermal fluctuations.
We add to the effective field in
Eq.~(\ref{eq:landaulifschitzgilbert}) stochastic contributions
$\mathbf{h}$ such that
$\mathbf{H}\rightarrow\mathbf{H}+\mathbf{h}$, where $\mathbf{h}$
has white-noise correlations
\begin{align}\label{eq:stochasticcorrelations}
\langle h_i(\vec{x},t) h_j(\vec{x}',t')\rangle=&\sigma_{ij}
\delta(\vec{x}-\vec{x}')\delta(t-t')\;,\\\nonumber\\
\langle h_i(\vec{x},t)&\rangle=0\;,
\end{align}
and where the indices $i,j\in\{x,y,z\}$ label Cartesian
coordinates. The strength $\sigma_{ij}$ is given by the
fluctuation-dissipation theorem $\sigma_{ij}=\delta_{ij}2\alpha\kB
Ta^3/\hbar$, which assures that, in the absence of field and
current, the Boltzmann equilibrium distribution
\begin{equation}\label{eq:boltzmannequilibrium}
P_{\rm eq}[\mathbf{\Omega}]\propto e^{-E_{\rm
MM}[\mathbf{\Omega}]/\kB T}\;,
\end{equation}
is reached after sufficiently long times. In principle,
Eqs.~(\ref{eq:eulerlagrange}--\ref{eq:stochasticcorrelations})
describe the full magnetization dynamics. Obtaining results on
finite-temperature average drift velocities of domain walls is
however very cumbersome, especially in the presence of extrinsic
pinning. We therefore use a variational method.

We obtain the specific form of our variational {\rm ansatz} by
varying the action in Eq.~(\ref{eq:actionint}) for a
time-independent magnetization. Using
$\om=(\sin\theta\cos\phi,\sin\theta\sin\phi,\cos\theta)$ we find
$\theta''=(K_z/J)\sin^2\theta$,\cite{tatara2004} where the primes
denote derivatives with respect to $x$. This equation has
domain-wall solutions $\tan(\theta/2)=\exp\{\pm[x-X]/\lambda\}$
and $\phi\in\{0,\pi\}$, where $\lambda=\sqrt{J/K_z}$ is the
domain-wall width and $X$ is the position of the domain wall.

The variational {\it ansatz} we use is $\tan(\theta_{\rm
dw}/2)=\exp\{[x-X(z,t)]/\lambda\}$ and $\phi=\phi_0(z,t)$. A
domain wall is now described by two collective coordinates
$X(z,t)$ and $\phi_0(z,t)$ that represent the position of the
domain wall line and the chirality at this position, respectively.
Note that $z$ is the coordinate along the line, and also that
there are other possibilities for the exact form of the {\it
ansatz}, such as choosing a different domain-wall charge.

We choose the magnetic field pointing in the positive $z$
direction $\mathbf{B}_{\rm ext}=B_z\hat{z}$, $B_z>0$. Furthermore,
the current is taken in the positive $x$ direction. We find that
the action in terms of the collective coordinates $X(z,t)$ and
$\phi_0(z,t)$ is then given by
\begin{align}\label{eq:actiontot}
S[X,\phi_0]=&\nonumber\\
-N\hbar \int & d t\int \frac{d z}{L_z}\,
\bigg(\frac{X}{\lambda}\dot\phi_0
+\frac{K_\perp}{\hbar}\sin^2\phi_0 \nonumber\\
&+\frac{J}{\hbar}\frac{(X')^2}{\lambda^2}
+\frac{J}{\hbar}(\phi_0')^2 -\frac{gB_z}{\hbar}\frac{X}{\lambda}
+\frac{v_\s}{\lambda}\phi_0\bigg)\;,
\end{align}
where $N=2\lambda L_y L_z/a^3$ is the number of magnetic moments
in a domain wall, with $L_y, L_z$ the length of the sample in the
$y$ and $z$ direction, respectively. We now substitute the {\it
ansatz} in Eq.~(\ref{eq:dissipation}) to obtain the dissipation
functional as a function of the collective coordinates
\begin{align}\label{eq:dissipationansatz}
R[X,\phi_0]=N\frac{\alpha\hbar}{2}\int\! d t\int \!\frac{d z}{L_z}
\,\Bigg[\left(\frac{\dot{X}}{\lambda}-\frac{\beta v_{\rm s}}
{\alpha\lambda}\right)^2\!\!+\dot\phi_0
\phantom{\lefteqn{\phi}}^2\Bigg]\;.
\end{align}
Upon variation of the total action in Eq.~(\ref{eq:actiontot})
with respect to $X$ and $\phi_0$, and setting this equal to the
variation of the dissipation function in
Eq.~(\ref{eq:dissipationansatz}) with respect to $\dot{X}$ and
$\dot\phi_0$, respectively, we obtain equations of motion for the
collective coordinates
\begin{align}\label{eq:langevin1}
\dot\phi_0+\alpha\frac{\dot{X}}{\lambda}&=\frac{2J}{\hbar}\frac{X''}
{\lambda}+\beta\frac{v_{\rm s}}{\lambda}+\frac{gB_z}{\hbar}\;,\\
\nonumber\\\label{eq:langevin2}
\frac{\dot{X}}{\lambda}-\alpha\dot\phi_0=&-\frac{2J}{\hbar}\phi_0''
+\frac{K_\perp}{\hbar}\sin(2\phi_0) +\frac{v_{\rm s}}{\lambda}\;.
\end{align}

We now add thermal fluctuations that contribute as stochastic
terms to Eqs.~(\ref{eq:langevin1})~and~(\ref{eq:langevin2}), so
that we obtain Langevin equations
\begin{align}\label{eq:langevinnew1}
\dot\phi_0(z,t)+\alpha\frac{\dot{X}(z,t)}{\lambda}
=&-\frac{L_z\lambda}{\hbar N}\frac{\delta V_{\rm
eff}[X,\phi_0]}{\delta X(z)}
+\eta_X(z,t)\;,\\
\nonumber\\\label{eq:langevinnew2}
\frac{\dot{X}(z,t)}{\lambda}-\alpha\dot\phi_0(z,t)
&=\frac{L_z}{\hbar N}\frac{\delta V_{\rm eff}[X,\phi_0]}{\delta
\phi_0(z)}+\eta_{\phi_0}(z,t)\;.
\end{align}
Here we take functional derivatives of an effective potential that
is a functional of $X(z)$ and $\phi_0(z)$. Allowing for an
arbitrary potential $V_{\rm dis}(X,\phi_0)$ due to disorder and
inhomogeneities, this effective potential is given by
\begin{align}\label{eq:Vtotal}
V_{\rm eff}&[X,\phi_0]=\nonumber\\&-\hbar
N\int_0^{L_z}\frac{dz}{L_z}\bigg[-\frac{J}{\hbar}\left(\frac{X'^2}{\lambda^2}+\phi_0'^2\right)
+\frac{K_\perp}{2\hbar}\cos(2\phi_0) \nonumber\\&+\frac{v_{\rm
s}}{\lambda}\left(\beta \frac{X}{\lambda}-\phi_0\right)\!
+\frac{gB_z}{\hbar} \frac{X}{\lambda}\bigg]+ V_{\rm
dis}(X,\phi_0)\;,
\end{align}
and also includes contributions from the micromagnetic energy
functional, the external field, and spin-transfer torques. Note
that in the absence of current, field and disorder, the potential
in Eq.~(\ref{eq:Vtotal}) is exactly the total micromagnetic energy
$E_{\rm MM}[\om_{\rm dw}]=V_{\rm eff}[X,\phi_0]$.

The noise in
Eqs.~(\ref{eq:langevinnew1})~and~(\ref{eq:langevinnew2}) obeys
$\langle\eta_i(z,t)\eta_j(z',t')\rangle=\sigma\delta_{ij}\delta(t-t')\delta(z-z')$
and $\langle\eta_i(z,t)\rangle=0$, where $\{i,j\}\in\{X,\phi_0\}$.
The strength $\sigma$ can be determined from the Fokker-Planck
equation, which for the Langevin
equations~(\ref{eq:langevinnew1})~and~(\ref{eq:langevinnew2}) is
given by\cite{riskenbook}
\begin{align}\label{eq:fokkerplanck}
(1+\alpha^2)&\frac{\partial P[X,\phi_0]}{\partial
t}=\nonumber\\
\frac{1}{\hbar}\int_0^{L_z}\!&\frac{d
z}{L_z}\bigg\{\frac{\delta}{\delta
\phi_0}\bigg[\bigg(\frac{\alpha}{N}\frac{\delta V_{\rm
eff}}{\delta \phi_0}+\frac{\lambda}{N}\frac{\delta V_{\rm
eff}}{\delta
X}\bigg) P[X,\phi_0]\bigg]\nonumber\\
&\;+\lambda\frac{\delta}{\delta X}\bigg[\bigg(\frac{\alpha\lambda
}{N}\frac{\delta V_{\rm eff}}{\delta X}-\frac{1}{N}\frac{\delta
V_{\rm eff}}{\delta\phi_0}\bigg)
P[X,\phi_0]\bigg]\nonumber\\
&\;+\frac{\sigma}{2}\bigg(\lambda^2\frac{\delta^2}{\delta
X^2}+\frac{\delta^2}{\delta \phi_0^2}\bigg)P[X,\phi_0]\bigg\}\;.
\end{align}
By demanding that the equilibrium Boltzmann distribution function
that follows from Eq.~(\ref{eq:boltzmannequilibrium}), given by
\begin{align}\label{eq:equilibriumdistribution}
P_{\rm eq}\propto e^{-V_{\rm eff}/\kB T}\;,
\end{align}
is a time-independent solution of the above Fokker-Planck
equation, we find the strength of the thermal fluctuations as
\begin{align}\label{eq:sigmaline}
\sigma=2\alpha \kB T L_z/\hbar N\;.
\end{align}
We see that the noise obeys the fluctuation-dissipation theorem
with an effective temperature $T L_z/N$. The temperature is
therefore effectively reduced by the magnetic-moment density in
the domain wall line.

\subsection{Results}

Two of us analyzed the model in Eqs.
(\ref{eq:langevinnew1})~and~(\ref{eq:langevinnew2}) in the
presence of extrinsic pinning.\cite{duine2008} In this section, we
focus instead on the clean situation, in which $V_{\rm dis} = 0$
everywhere. In the $T=0$ case, the domain wall line will stay
straight because the force on each point is exactly the same.
Specializing to $\beta = 0$, we find that there is a critical
current $v_{\rm s, crit} = \lambda K_\perp/\hbar$. Below this
critical current, the domain wall will not be able to acquire a
finite drift velocity. This phenomenon is usually called intrinsic
pinning,\cite{tatara2004} and it does not occur for $\beta \neq
0$. Above the critical current, the domain wall will acquire an
average drift velocity, given by
\begin{align}
\langle \dot{X} \rangle = -\frac{1}{(1 + \alpha^2)} \sqrt{v_\s^2 -
v_{\rm s, crit}^2 }\;.\label{eq:xdot}
\end{align}
In the presence of thermal fluctuations we can no longer assume
that the domain wall remains straight, and we need to go through a
rather more elaborate procedure to find the average drift
velocity. More specifically, if there are thermal fluctuations, we
can differentiate between the \emph{flow} regime above $v_{\rm s,
crit}$, for which Eq.~(\ref{eq:xdot}) still approximately applies,
and the \emph{thermal} regime below $v_{\rm s, crit}$, in which
the speed is finite, but goes with a different power law.

To find the behavior in the thermally assisted regime, we start by
rewriting the Langevin equations to find just one equation for
$\phi_0(z, t)$. We then specialize to the case without an external
magnetic field, with $\beta=0$, and take $1 + \alpha^2 \simeq 1$.
We assume that $X''\simeq 0$ because there is no potential that
couples to $X$ and find

\begin{align}
\dot{\phi_0} =2 \alpha \frac{J}{\hbar}\phi_0'' - \frac{\alpha
K_\perp}{\hbar} \sin(2 \phi_0) - \frac{\alpha v_\s}{\lambda} +
\eta_X - \alpha \eta_{\phi_0}\, ,
\end{align}
which describes the motion of a string in a tilted washboard
potential, a problem that was investigated before in a different
context by B\"{u}ttiker and Landauer.\cite{buttiker1981} For $v_\s
< v_{\rm s, crit}$, the string propagates by thermal activation.
This occurs due to the formation of a nucleus, or of a
kink-antikink pair in the string. That is, part of the string is
moved over the potential barrier due to thermal activation. If
this nucleus is large enough, the kink and antikink will proceed
to move apart from one another, and the string propagates to the
next potential valley. Two factors are important: the energy
barrier $\Delta E$ that needs to be overcome to generate a
sufficiently large nucleus, and the propagation velocity of the
kinks (and antikinks).

\begin{figure}[h!]
\begin{center}
\includegraphics[width=8cm]{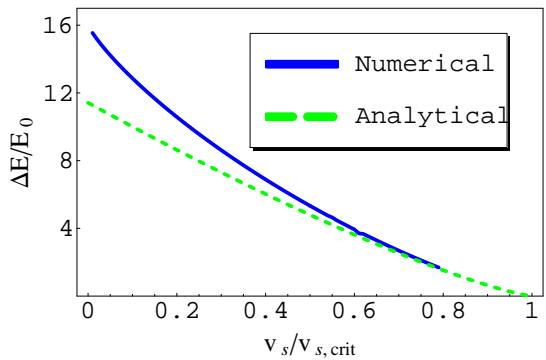}\\
\caption{(Color online) The energy barrier $\Delta E/E_0$
necessary to generate a nucleus large enough for the kink and
antikink to propagate as a function of $v_\s/v_{\rm s, crit}$. The
solid line was found by numerical calculation. The dashed line
corresponds to Eq.~(\ref{eq:DE}).} \label{fig:DE}
\end{center}
\end{figure}

Let us start with the former. It is given by $\Delta E/ E_0=
\sqrt{J/K_\perp}\int (d\phi_{0, N}/dz)^2 dz$, where $\phi_{0, N}$
is a time-independent solution of the differential equation
$(J/K_\perp)(d^2 \phi_{0, N}/d z^2) = \sin(\phi_{0, N}) -
v_\s/v_{\rm s, crit}$, and represents a stationary configuration,
corresponding to the motion from a local maximum and back. In the
above, $\Delta E$ is given in units of $E_0 = \alpha \sqrt{J
K_\perp}/(2\lambda)$. The result is shown in Fig.~\ref{fig:DE}.
In the limit that we are close to the critical current, i.e. that
$(v_{\rm s, crit} - v_\s)/v_{\rm s, crit} \ll 1$, we can solve for
$\Delta E$ exactly and find\cite{buttiker1981}

\begin{align}\label{eq:DE}
\Delta E = \frac{24}{5} \alpha\frac{\sqrt{J K_\perp}}{\lambda}
\left(1 - \frac{v_\s}{v_{\rm s, crit}} \right)^{5/4}\, .
\end{align}
This formula indeed fits very well to the tail of our numerical
curve. Note that the energy barrier remains finite as $v_\s
\rightarrow 0$. The difference between the above formula and the
numerical solution is at most 25\%, suggesting we may use
Eq.~(\ref{eq:DE}) to estimate the qualitative behavior of domain
wall motion even at lower values of $v_\s/v_{\rm s, crit}$. In the
limit that $v_\s\rightarrow 0$ we have that $\Delta E\propto
v_\s\log v_\s$. This limit is not shown in Fig.~\ref{fig:DE} as it
applies only for $v_\s$ very close to zero.

\begin{figure}[h!]
\begin{center}
\includegraphics[width=7.5cm]{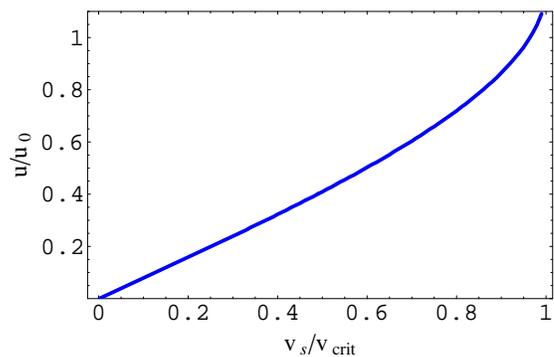}\\
\caption{Velocity of the kink (equal to minus the velocity of the
antikink), as a function of current.}\label{fig:u}
\end{center}
\end{figure}

The other important quantity is the velocity at which the kink and
antikink move away from one another. This velocity is found
numerically by solving the equation
\begin{align}
\phi_0'' + \frac{u}{u_0} \phi_0 ' - \sin\phi_0 +
\frac{v_\s}{v_{\rm s, crit}} = 0\, ,
\end{align}
with $u_0=2\lambda E_0/\hbar$, and finding the $u$ for which the
solution $\phi_0(z)$ which starts out at $\phi(0) =
\arcsin(v_\s/v_{\rm s, crit})$ will go away from that point, and
return to $\arcsin(v_\s/v_{\rm s, crit})$ at $z \rightarrow
\infty$. The universal curve for this velocity is shown in
Fig~\ref{fig:u}.

The probability of creating sufficiently large nuclei follows an
Arrhenius law $j \propto \exp [- \Delta E/\kB T)]$. We now have
all the necessary ingredients to find the average velocity of the
string, which is proportional to $\sqrt{u j}$. In the limiting
case where the current is close to the critical one, we have
that\cite{buttiker1981}

\begin{figure}[h!]
\begin{center}
\includegraphics[width=8.3cm]{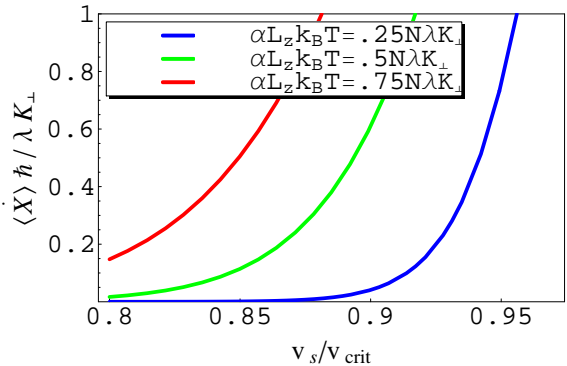}\\
\caption{(Color online) Velocity of the domain wall line as a
function of current, for different values of
$T$.}\label{fig:speeds}
\end{center}
\end{figure}

\begin{widetext}\begin{align}
\langle\dot{X}\rangle =  5^{3/4} (6 \pi)^{1/4} \frac{2\lambda
K_\perp}{\hbar}  \sqrt{u}\;e^{-\frac{\Delta E}{2\kB T }}
\left(\frac{\Delta E }{\kB T }\right)^{1/4}\left[1 -
\left(\frac{v_\s}{v_{\rm s, crit}}\right)^2\right]^{3/8} \;,
\end{align}\end{widetext}
where $u$ is a function of $v_\s/v_{\rm s, crit}$ as in
Fig.~\ref{fig:u}. In Fig.~\ref{fig:speeds}, we have plotted this
velocity for different temperatures.

\subsection{Experimental Status}

In many experiments, the nanostrip is sufficiently narrow that we
can neglect deformations of the domain wall line and approximate
it as being rigid, an approximation we treat in the next section.
However, Yamanouchi {\it et al.}\cite{yamanouchi2007} have
observed in their experiments with magnetic semiconductors that
the domain wall looks wedge-shaped in the current-induced case,
suggesting that deformations play a role in wide enough
nanostrips.

Yamanouchi {\it et al.} also found that the velocity of the domain
wall obeys a scaling law. More specifically, they fitted their
data with a creep-like scaling law $\log\dot{X}\propto
v_\s^{-\mu}$ with an exponent $\mu \simeq 0.33$. Recent
experiments with ferromagnetic metals\cite{moore2008} have found
the exponent $\mu\simeq1/4$ which would imply\cite{duine2008} that
the dissipative spin-transfer torque dominates in the creep regime
in this case.

\section{Rigid domain walls}\label{section:rigid domain walls}

In this section, we simplify the model of domain-wall lines by
assuming that the domain-wall coordinates are constant along the
$z$ direction, the domain wall is then rigid. This simplification
allows us to obtain various results analytically.

\subsection{Model}

As mentioned before, rigid domain walls obey $X'=\phi_0'=0$, i.e.,
they are rigid in the $z$ direction. We expect this approximation
to hold in the limit when $L_z$ is comparable to $\lambda$. We
integrate Eqs.~(\ref{eq:langevinnew1})~and~(\ref{eq:langevinnew2})
over $z$ (which is quite trivial since there are no $z$
dependences anymore) to obtain the Langevin equations for a rigid
domain wall (We use $\int dz \delta V[X(z)]/\delta
X(z)\rightarrow\partial V(X)/\partial X$ and equivalently for
$\phi_0$)
\begin{align}\label{eq:langevinrigid1}
\dot\phi_0+\alpha\frac{\dot{X}}{\lambda}&= \frac{-\lambda}{\hbar N
}\frac{\partial V_{\rm rigid}}{\partial X
}+\tilde\eta_X\;,\\\nonumber\\\label{eq:langevinrigid2}
\frac{\dot{X}}{\lambda}-\alpha\dot\phi_0&=\frac{1}{\hbar
N}\frac{\partial V_{\rm
rigid}}{\partial\phi_0}+\tilde\eta_{\phi_0}\;.
\end{align}
We integrate the total potential in Eq.~(\ref{eq:Vtotal}) to find
(note that $X'=\phi_0'=0$)
\begin{align}\label{eq:Vrigid}
V_{\rm rigid}=-\hbar
N&\bigg[\frac{K_\perp}{2\hbar}\cos(2\phi_0)+\frac{gB_z}{\hbar}\frac{X}{\lambda}\nonumber\\
&+\frac{v_\s}{\lambda}\left(\beta\frac{X}{\lambda}
-\phi_0\right)\bigg]+V_{\rm dis}(X,\phi_0)\;.
\end{align}
The stochastic correlations are found from
\begin{align}
\langle\tilde\eta_i(t)\tilde\eta_j(t')\rangle
=\frac{1}{L_z^2}\langle\int_0^{L_z}dz&\int_0^{L_z}dz'\eta_i(z,t)\eta_j(z',t')\rangle
\nonumber\\&=\frac{2\alpha\kB T}{\hbar
N}\delta_{ij}\delta(t-t')\;.
\end{align}
In our model, rigid domain-walls obey the fluctuation-dissipation
theorem with effective temperature $T/N$, i.e., the temperature is
effectively reduced by the number of magnetic moments in the
domain wall.

\subsection{Results}

\subsubsection{Clean system, intrinsic pinning}

We first focus on the case that the extrinsic pinning is zero.
Substitution of $\dot{X}$ from Eq.~(\ref{eq:langevinrigid1}) into
Eq.~(\ref{eq:langevinrigid2}) then provides us with an equation
that is independent of $X$. Using the equilibrium solution of
Eq.~(\ref{eq:fokkerplanck}), we find the average velocity of the
chirality $\langle\dot{\phi}_0\rangle$. With this result and
Eq.~(\ref{eq:langevinrigid1}), we compute average drift velocities
\begin{equation}\label{eq:driftvelocity}
\alpha\frac{\langle\dot{X}\rangle}{\lambda}=
-\langle\dot{\phi}_0\rangle+\beta\frac{v_\s}{\lambda}+\frac{gB_z}{\hbar}\;,
\end{equation}
where the average chirality velocity is given by\cite{riskenbook}
(we omit a factor $1+\alpha^2\simeq1$)
\begin{align}\label{eq:driftvelocityangle}
&\langle\dot{\phi}_0\rangle=\nonumber\\
&\frac{-2\pi(e^{H_{\rm eff}}\!-\!1)\alpha\kB T/\hbar N}{
\int_0^{2\pi} d \phi e^{-\Phi(\phi)} \Bigg[\int_0^{2\pi}d \phi'
e^{\Phi(\phi')}+(e^{H_{\rm eff}}\!-\!1)\int_{0}^{\phi}
d\phi'e^{\Phi(\phi')}\Bigg]}.
\end{align}
In this expression, the dimensionless effective potential is given
by
\begin{align}
\Phi(\phi_0)=\frac{N\hbar}{\kB T}\left[\frac{\sigma H_{\rm
eff}}{4\pi L_z}\frac{\phi_0}{\alpha}
-\frac{K_\perp}{2\hbar}\cos(2\phi_0)\right]\;,
\end{align}
and the dimensionless effective field is defined as $H_{\rm
eff}=4\pi L_z[(\alpha-\beta)v_\s/\lambda-gB_z/\hbar]/\sigma$. The
expressions in
Eqs.~(\ref{eq:driftvelocity})~and~(\ref{eq:driftvelocityangle})
generalize the results by Duine {\it et al.},\cite{duine2007b} to
include external magnetic fields and $\beta\neq 0$.

\begin{figure}[h!]\centering
\includegraphics[width=8.5cm]{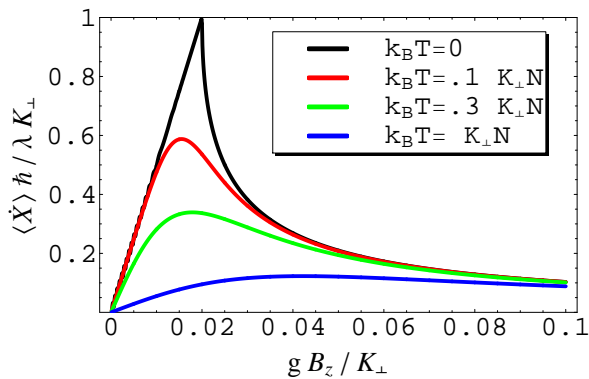}
\caption{(Color online) Field-driven domain-wall motion for
several temperatures. We set $v_\s=0$ and take
$\alpha=0.02$.}\label{fig:fielddrivendrift}
\end{figure}

In the field-driven case, we set $v_{\rm s}=0$ to find the
behavior in Fig.~\ref{fig:fielddrivendrift}. In the calculations,
we use a fixed value for the damping parameter $\alpha=0.02$ and
several values for the temperature.
At zero temperature, the drift velocity depends on the external
magnetic field linearly as $\langle\dot{X}\rangle=\lambda
(gB_z/\hbar)/\alpha$ up to a critical value $H_{\rm cr}=\alpha
K_\perp/\hbar$. At this point, the domain wall starts precessing,
causing the Walker-breakdown. This behavior was originally
predicted by Schryer and Walker,\cite{walker1974} and was
subsequently observed, e.g. by Beach {\it et al.}\cite{beach2005}
From Fig.~\ref{fig:fielddrivendrift} we see that for nonzero
temperatures, the Walker breakdown smoothens out. For
$T\rightarrow\infty$, it fully disappears, and we find that the
domain-wall velocity is linear with the field for all fields. We
note that temperature only has an effect on the drift velocity for
small fields. For very large external fields, the drift velocity
is for all temperatures linear with the field and obeys
$\langle\dot{X}\rangle=\alpha\lambda (gB_z/\hbar)$.

For the purely current-driven case, where we set $B_z=0$, the
relative values of $\alpha$ and $\beta$ determine the sign of the
contribution due to the current to the effective force. Again, we
set $\alpha=0.02$ and choose several values for $\beta$ and for
the temperature.
\begin{figure}[h!]
\includegraphics[width=8cm]{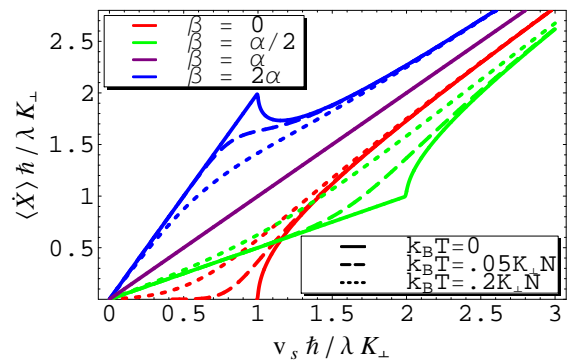}
\caption{(Color online) Current-driven domain-wall motion for
several temperatures and several values of $\beta$. We set $B_z=0$
and take $\alpha=0.02$.}\label{fig:currentdrivendrift}
\end{figure}
It is indeed seen in Fig.~\ref{fig:currentdrivendrift} that the
behavior of the average drift velocity for $\beta<\alpha$ is very
different from $\beta>\alpha$. In the limiting case $\beta=0$, we
see that there is a critical current $v_{\rm s,
crit}=K_\perp\lambda/\hbar$, in agreement with
Eq.~(\ref{eq:xdot}). In the large $\beta$ limit, we recognize a
Walker-breakdown-like behavior, just like the behavior found in
purely field-driven domain-wall motion. For very large currents,
the drift velocity acquires a linear dependence on the spin
current $\langle\dot{X}\rangle=v_\s$ for all values of $\beta$ and
for all temperatures.

\subsubsection{Extrinsic pinning}

Domain-wall pinning is of practical interest. Control of domain
walls is achieved by pinning the domain wall in a position by
means of, for example, a deformation in the material. In nanowires
or thin metal strips, dents in the sample act as an intended
pinning potential. A thorough understanding of depinning times is
especially important in systems that are relevant for
technological applications, such as data-storage devices.

To incorporate pinning in our theory, we add a pinning potential
$V_{\rm pin}$ to the potential $V_{\rm rigid}$ in
Eqs.~(\ref{eq:langevinrigid1})~and~(\ref{eq:langevinrigid2}). A
pinning potential that is due to a deformation in the material can
be obtained through energy analysis. We discern between two types
of deformations: symmetric and asymmetric notches. The sample we
have in mind is a thin strip of ferromagnetic or semiconductor
material, where the notches are little dents on the sides of the
strip. In the case of symmetric notches, there is a dent on both
sides of the sample. If there is only a dent on one side, we have
an asymmetric notch. It turns out that symmetric pinning sites can
effectively be described by a quadratic-well potential in the $X$
direction,\cite{hayashi2006} that is independent of the chirality.
For asymmetric notches, the pinning potential also has a chirality
dependence.\cite{petit2007}

As an example, we consider the symmetric notch. If we add a $X$
dependent symmetric-notch contribution to
Eqs.~(\ref{eq:langevinrigid1})~and~ (\ref{eq:langevinrigid2}), we
see that the pinning contribution in Eq.~
(\ref{eq:langevinrigid2}) vanishes. Because of the explicit $X$
dependence in Eq.~(\ref{eq:langevinrigid1}), the system can no
longer be described by a probability distribution that depends on
the variable $\phi_0$ only. Therefore, all terms in the
Fokker-Planck equation~(\ref{eq:fokkerplanck}) need to be taken
into account, and we are not able to find an analytic solution to
the full problem. We can, however, compute depinning times using
Kramer's escape-rate theory.

The potential that Tatara {\it et al.}\cite{tatara2004} proposed
for a symmetric notch has a kink at the sides, which makes it less
suitable for escape-rate computations. We therefore use a very
similar, but smooth potential
\begin{equation}\label{eq:symmetricpotential}
V_{\rm pin}^{\rm S}=\frac{N
V_0}{2}\left[2\left(\frac{X}{\xi}\right)^2
-\left(\frac{X}{\xi}\right)^4\right]\;,
\end{equation}
where $V_0/2$ is the depth and $2\xi$ is the width of the
potential well.

We compute depinning times for the pinning potential in
Eq.~(\ref{eq:symmetricpotential}) using Kramer's escape-rate
theory which states that the depinning time is proportional to
$\log(\tau\omega_0)=\Delta V/\hbar\kB T$, where $\omega_0\propto
V_0/\hbar$ is an attempt frequency and $\Delta V$ is the height of
the potential barrier that has to be overcome. We determine the
positions of the potential minimum and the saddle through
variation of the potential $V_{\rm rigid}+V_{\rm pin}$.
Substitution of these coordinates provides us with the potential
difference $\Delta V$. The depinning time for driving current and
field is found as $\log{\tau}\propto \Delta V\propto
-(gB_z/\hbar+\beta v_\s/\lambda)$. Note that if $\beta=0$, the
depinning time is independent of the applied current within this
approximation.

\subsubsection{Disorder potential}

The final potential that we consider is a disorder potential. Due
to, for example, roughness on the edges of a sample, there is a
random pinning potential $V_{\rm dis}(X)$ that is felt by the
domain wall. We consider a situation of strong pinning of the
angle $\phi_0$ such that $\langle\dot\phi_0\rangle=0$ (See
Ref.~[\onlinecite{lecomte2009}] for an extensive analysis
including the dynamics of this angle). This would be the case for
fields below Walker breakdown and currents below $v_{\rm s,
crit}$. The equation of motion for the coordinate $X$ is then
found from Eq.~(\ref{eq:langevinrigid1}) to be
\begin{align}
\alpha\frac{\dot{X}}{\lambda}=\beta\frac{v_\s}{\lambda}+\frac{gB_z}{\hbar}
-\frac{\lambda}{\hbar N}\frac{\partial V_{\rm dis}(X)}{\partial
X}+\tilde\eta_X\;.
\end{align}
The disorder potential $V_{\rm dis}(X)$ that enters the equations
of motion is characterized by certain spatial correlations
$\overline{\;[V_{\rm dis}(X)-V_{\rm
dis}(X')]^2}=\Delta|(X-X')/\lambda|^\gamma$. In this expression,
the line denotes an average over the disorder, and $\Delta>0$ is a
measure for the strength of the disorder potential. The exponent
$\gamma$ characterizes the nature of the correlations. In general,
there are two ways of obtaining a disorder potential: by applying
a random field, or by randomizing locally the strength of certain
coupling constants in the system, for example the anisotropy or
the spin stiffness. The former is called random field disorder and
gives rise to correlations with $\gamma=1$, whereas the latter is
called random bond disorder with exponent $\gamma=0$. For both
limits, Le Doussal and Vinokur\cite{ledoussal1995} have obtained
expressions for depinning times for a zero-dimensional object (in
our one-dimensional model, the domain wall itself is described as
a point at position $\{X,\phi_0\}$ with dimension zero).

For random field disorder ($\gamma=1$), Le Doussal and Vinokur
find that the drift velocity is zero up to some critical driving
current and/or field $F_{\rm c}\propto\Delta/T$, and is then
linear with the driving force
\begin{equation}
\frac{\langle\dot{X}\rangle}{\lambda}=\frac{\beta
v_\s}{\lambda}+\frac{gB_z}{\hbar}-F_{\rm c}\;.
\end{equation}
For random bond disorder
($\gamma=0$) they find that the drift velocity obeys
\begin{equation}
\langle\dot{X}\rangle\propto\frac{(\beta
v_\s/\lambda+gB_z/\hbar)^{z-1}}{\Gamma(z-1)T^{z+2}}\;,
\end{equation}
where $z=2+\Delta/2(\kB T)^2$ is the so-called dynamical exponent
and $\Gamma(z)$ is the gamma function. Surprisingly, they find
that for $0<\gamma<1$ the results turn out to resemble the results
obtained for higher-dimensional objects, i.e. Le Doussal and
Vinokur find a creep scaling law with a certain creep exponent
\begin{align}
\langle\dot{X}\rangle&\propto \left(\frac{\beta
v_\s/\lambda+gB_z/\hbar
}{F_{\rm c}}\right)^{\frac{z-1-\gamma/2}{1-\gamma}}\times\nonumber\\
&\exp\left[-(1-\gamma)\left( \frac{\beta v_\s/\lambda+gB_z/\hbar
}{\gamma F_{\rm c}}\right)^{\frac{-\gamma}{1-\gamma}}\right]\;,
\end{align}
where $F_{\rm c}\propto T[\Delta/2(\kB T)^2]^{1/\gamma}$ is a
characteristic critical driving force. Note that the creep
exponent $(z-1-\gamma/2)/(1-\gamma)$ is always larger than $1$ for
$0<\gamma<1$.

\subsection{Comparison with other work}

To compute drift velocities at finite temperature, we have
expanded the theory proposed by Duine {\it et
al.}\cite{duine2007a} to include external magnetic fields in
addition to driving currents. Other theoretical work is done by
Tatara {\it et al.}\cite{tatara2005} and by Martinez {\it et
al.}\cite{martinez2007} Here, we compare our results with the
results by Martinez {\it et al.}

In order to write
Eqs.~(\ref{eq:langevinrigid1})~and~(\ref{eq:langevinrigid2}) in
terms of the coordinate $X$ only, Martinez {\it et al.} assume
$\phi_0$ to be small such that $\sin2\phi_0\simeq2\phi_0$. Note
that this assumption only holds when $B_z\ll g K_\perp$ and/or
$v_\s\ll v_{\rm s,crit}$ for $\beta\neq 0$. If, however, we do
make this assumption, and differentiate
Eqs.~(\ref{eq:langevinrigid1})~and~(\ref{eq:langevinrigid2}) with
respect to time, we find
\begin{align}\label{eq:langevinrigidcolor1}
\frac{\ddot{X}}{\lambda}=\alpha\ddot\phi_0&+2\dot\phi_0\frac{K_\perp}{\hbar}\;,
\\\nonumber\\\label{eq:langevinrigidcolor2}
\ddot\phi_0=&-\alpha\frac{\ddot{X}}{\lambda}\;,
\end{align}
where we omitted the stochastic terms. We now substitute
Eqs.~(\ref{eq:langevinrigid1})~and~(\ref{eq:langevinrigidcolor2})
in Eq.~(\ref{eq:langevinrigidcolor1}) and like Martinez {\it et
al.} add a stochastic term to find
\begin{align}\label{eq:langevinrigidcolor3}
(1+\alpha^2)\frac{\hbar}{2K_\perp}\frac{\ddot{X}}{\lambda}=
-\alpha\frac{\dot{X}}{\lambda}+\beta\frac{v_\s}{\lambda}+\frac{gB_z}{\hbar}+\zeta_{\rm
th}\;.
\end{align}
The correlations of the stochastic force are, according to
Martinez {\it et al.}, given by
\begin{align}
\langle \zeta_{\rm th}(t)\zeta_{\rm
th}(t')\rangle=\sigma\delta(t-t')\;,\qquad\langle \zeta_{\rm
th}(t)\rangle=0\;.
\end{align}
From the fluctuation-dissipation theorem, they infer that
$\sigma\propto N\alpha\kB T$. An easy way to obtain the
Fokker-Planck equation is to introduce a new variable
$v=\dot{X}$,\cite{riskenbook} such that we have two Langevin
equations
\begin{align}\label{eq:langevinrigidcolor4}
\dot{X}=&v\;,
\\\nonumber\\\label{eq:langevinrigidcolor5}
(1+\alpha^2)\frac{\hbar}{2K_\perp}\frac{\dot{v}}{\lambda}
=-\alpha\frac{v}{\lambda}&+\beta\frac{v_\s}{\lambda}+\frac{g B_z
}{\hbar}+\zeta_{\rm th}\;.
\end{align}
Note that the above equations have to be solved with the initial
conditions $\dot{X}(t=0)=v(t=0)=v_\s$ to include the reactive
spin-transfer torque. The Fokker-Planck equation generated by
these Langevin equations is given by (we omit a factor
$1+\alpha^2\simeq1$)
\begin{align}\label{eq:fokkerplanckcolorednoise}
&\frac{\partial P[X,v]}{\partial t}=-\frac{\partial}{\partial X}(v
P[X,v])\nonumber\\
&-\frac{2K_\perp}{\hbar}\frac{\partial}{\partial v}\bigg[
-\alpha\frac{v}{\lambda}+\beta\frac{v_\s}{\lambda}+\frac{g B_z
}{\hbar}+\frac{\sigma K_\perp}{\hbar}\frac{\partial}{\partial v}
\bigg]P[X,v]\;.
\end{align}
This Fokker-Planck equation is satisfied by a Boltzmann
equilibrium distribution that has the same potential energy as
Eq.~(\ref{eq:Vrigid}) for $\phi_0=0$, but with an additional
kinetic energy $\hbar v^2/K_\perp$. The exact form of the
stochastic strength is $\sigma=\alpha \kB T/NK_\perp$. We conclude
that this procedure complies with our model for a small range of
applicability.

Several field-driven domain-wall motion experiments have been
performed in ferromagnetic metallic
materials.\cite{beach2005,hayashi2006} Clear Walker-breakdown
behavior is observed, but the peak is not smoothed like our theory
predictions. Estimates by Duine {\it et al.}\cite{duine2007b} show
that room temperature, at which these experiments were performed,
leads to an effective temperature $\kB T/NK_\perp\simeq 10^{-3}$
for ferromagnetic metals. Our prediction is therefore
indistinguishable from the zero-temperature curve in
Fig.~\ref{fig:fielddrivendrift}. The reason for this low effective
temperature is the fact that the number of particles in the domain
wall is relatively high in a ferromagnetic metal. In magnetic
semiconductors, however, not all magnetic moments participate in
the magnetization, reducing the number of magnetic moments in a
domain wall by up to a factor $\sim 100$, thereby greatly
increasing the effective temperature. From
Fig.~\ref{fig:fielddrivendrift} we see that an effective
temperature $\kB T/NK_\perp\simeq 10^{-1}$ should be
distinguishable from the zero-temperature curve. We predict
therefore that the influence of thermal effects on Walker
breakdown should be observable with field-driven domain walls in
clean magnetic semiconductors.

Escape time studies have been performed on narrow domain walls by
Ravelosana {\it et al.}\cite{ravelosona2005} They indeed find that
the logarithm of the (average) escape time decreases linearly with
the applied current. Since the current dependence of the logarithm
of the escape time is determined by $\beta$, we can estimate the
value of $\beta$ from their curve, and we find that it is of the
order $\beta\sim 10^{-2}$, i.e. of the order of the damping
parameter $\alpha$, in agreement with theoretical expectations.

\section{Vortex domain walls}\label{section:vortex domain walls}

In this section we consider the limit of the domain-wall line that
corresponds to vortex domain walls and derive several analytical
and numerical results.

\subsection{Model}
Vortex domain walls are described by making the {\it ansatz} for
the $z$ dependence of the chirality
\begin{align}\label{eq:ansatzvortexfull}
\phi_0=2\arctan e^{\pm[z-Z(t)]/\kappa}\;.
\end{align}
The coordinate $Z(t)$ plays a similar role as $X(t)$, and the
width of the vortex domain wall $\kappa=\sqrt{J/K_\perp}$ in the
$z$ direction is the equivalent of the width $\lambda$ in the $x$
direction. The coordinates $\{X(t),Z(t)\}$ now determine the
position of the vortex at time $t$.

In principle, we should consider boundary conditions in the $z$
direction. We deal with this problem by assuming
$\kappa/L_z\rightarrow 0$, in which limit the {\it ansatz} reduces
to $\phi_0\simeq\pi/2\pm\{\theta[z-Z(t)]-1/2\}\pi$, with
$\theta(x)$ the Heaviside step function. The drawback is that we
are now neglecting all boundary effects. The $\pm$ sign is the
product of the chirality and charge in the {\it ansatz} and
determines whether we have a clockwise or counterclockwise
rotation in the vortex domain wall, and is usually denoted as the
Skyrmion number $s\in\{-1,1\}$ . We note that the vortex domain
wall is now fully characterized by its position $\{X(t),Z(t)\}$,
its dimensions $(\kappa\times\lambda)$, this Skyrmion number $s$
and the number of magnetic moments in the vortex $4\kappa\lambda
L_y/a^3=2\kappa N/L_z$.

We apply the simplified {\it ansatz} to the action in
Eq.~(\ref{eq:actiontot}) to find an action for a vortex domain
wall in terms of the collective coordinates $X(t)$ and $Z(t)$
\begin{align}\label{eq:actionvortex}
S_{\rm vor}[X,Z]=\nonumber\\
-\hbar N\int&\d
t\,\bigg(s\pi\frac{Z}{L_z}\frac{\dot{X}}{\lambda}-\frac{gB_z}{\hbar}\frac{X}{\lambda}
-s\pi\frac{v_\s}{\lambda}\frac{Z}{L_z}\bigg)\;.
\end{align}
The dissipation function in Eq.~(\ref{eq:dissipationansatz}) is
also written in terms of the new coordinates, however, we need to
take into account the full ansatz in
Eq.~(\ref{eq:ansatzvortexfull}) in order to find the $\kappa$
dependence in the last term
\begin{align}\label{eq:dissipationvortex}
R_{\rm vor}[X,Z]=\frac{\alpha\hbar N}{2}\int\d
t\,\Bigg[\left(\frac{\dot{X}}{\lambda}-\frac{\beta v_{\rm s}}
{\alpha\lambda}\right)^2+\frac{2\dot{Z}^2}{\kappa L_z}\Bigg]\;.
\end{align}
Variation of the functionals in
Eqs.~(\ref{eq:actionvortex})~and~(\ref{eq:dissipationvortex})
provides us with the Langevin equations for a vortex domain wall
\begin{align}\label{eq:langevinvortex1}
-s\pi\frac{\dot{Z}}{\kappa}+\frac{\alpha L_z
}{\kappa}\frac{\dot{X}}{\lambda}=\frac{\beta L_z
}{\kappa}&\frac{v_{\rm s}}{\lambda}+\frac{L_z
}{\kappa}\frac{gB_z}{\hbar}+\eta_X^{\rm
V}\;,\\\nonumber\\\label{eq:langevinvortex2}
s\pi\frac{\dot{X}}{\lambda}+2\alpha\frac{\dot{Z}}{\kappa}&
=s\pi\frac{v_{\rm s}}{\lambda}+\eta_Z^{\rm V}\;,
\end{align}
where we have again added stochastic forces to model thermal
effects. We see that the effective dampings are given by
$\alpha_X=\alpha L_z/\kappa$ and $\alpha_Z=2\alpha$. We write the
right-hand side of
Eqs.~(\ref{eq:langevinvortex1})~and~(\ref{eq:langevinvortex2}) in
terms of a total potential
\begin{align}
V_{\rm vortex} =-\hbar N\bigg[\frac{v_{\rm
s}}{\lambda}\left(\beta\frac{X}{\lambda}+s \pi\frac{Z}{L_z}\right)
+\frac{g B_z}{\hbar} \frac{X}{\lambda}\bigg]\;,
\end{align}
Note that this potential is also obtained by inserting the {\it
ansatz} in Eq.~(\ref{eq:Vtotal}). With these identifications, we
can again write the Langevin
equations~(\ref{eq:langevinvortex1})~and~(\ref{eq:langevinvortex2})
in the more suggestive form
\begin{align}\label{eq:langevinvortex1b}
s\pi\frac{\dot{Z}}{\kappa}-\alpha_X
\frac{\dot{X}}{\lambda}&=\frac{L_z}{\kappa
N}\frac{\lambda}{\hbar}\frac{\partial V_{\rm vortex}}{\partial X}
+ \eta_X^{\rm V}\;,\\\nonumber\\\label{eq:langevinvortex2b}
s\pi\frac{\dot{X}}{\lambda}+\alpha_Z\frac{\dot{Z}}{\kappa}&=
-\frac{L_z}{\kappa N}\frac{\kappa}{\hbar}\frac{\partial V_{\rm
vortex}}{\partial Z}+ \eta_Z^{\rm V}\;.
\end{align}
Note that the prefactor $L_z/N\kappa$ is proportional to the
inverse of the number of magnetic moments in the vortex domain
wall. Using the Fokker-Planck method outlined in section II, we
find that the probability distribution function $P$ does not
satisfy Boltzmann equilibrium $P\propto\exp[-V_{\rm
vortex}/\hbar\kB T]$ if we assume that the fluctuations in the $X$
and $Z$ direction have the same strength. However, we write down a
more general Fokker-Planck equation than the one in
Eq.~(\ref{eq:fokkerplanck}), in terms of stochastic correlations
$\langle\eta_i^{\rm V}(t)\eta_j^{\rm
V}(t')\rangle=\sigma_{ij}\delta(t-t')$. Again, we demand the
Boltzmann equilibrium in Eq~(\ref{eq:equilibriumdistribution}) to
be a solution to the modified Fokker-Planck equation, which yields
complicated equations that can be solved. Up to second order in
the small parameter $\alpha$, we find that the stochastic
correlations must obey
\begin{align}
\langle\eta_i^{\rm V}(t)\eta_j^{\rm
V}(t')\rangle=\sigma_i\delta_{ij}\delta(t-t')\;,
\end{align}
where the indices denote $i\in\{X,Z\}$ and
\begin{align}
\sigma_X =2 \alpha_X\frac{ \kB T L_z}{\hbar N\kappa}\,;\qquad
\sigma_Z =2 \alpha_Z\frac{ \kB T L_z}{\hbar N\kappa}\;.
\end{align}
In these relations, we recognize the fluctuation-dissipation
theorem with effective temperature $T_{\rm eff}=T L_z/N\kappa$.
The reduction by the factor $\kappa N/L_z$ is caused by the fact
that the number of microscopic degrees of freedom is proportional
to this factor. Note that, because the damping is anisotropic, it
is also necessary to introduce anisotropy in the fluctuations.

From the form of
Eqs.~(\ref{eq:langevinvortex1b})~and~(\ref{eq:langevinvortex2b}),
it is clear that we reach the isotropic case for
$\alpha_X=\alpha_Z$, i.e. when we demand that $L_z=2\kappa$. Note
that then also $\sigma_X=\sigma_Z$ and that the reduction of the
effective temperature is now proportional to $N$. That this is
indeed the isotropic case is also seen from the fact that now the
coordinates $X/\lambda$ and $Z/\kappa$ are treated on equal
footing in the dissipation functional in
Eq.~(\ref{eq:dissipationvortex}).

If we also demand that $\lambda=\kappa$, our model describes
circular vortices, that furthermore occupy the entire width of the
strip. This case is similar to the zero-temperature results of
Shibata {\it et al.}\cite{shibata2006} on the current-induced
vortex displacement in a magnetic nanodisk, and their equations of
motion correspond to ours in the case that $\beta = 0$ and $B_z =
0$.

To include extrinsic pinning in our model, we again add a pinning
potential to the potential $V_{\rm vortex}$. As an example, we
will consider a circularly symmetric pinning potential, quadratic
in both $X$ and $Z$, and bounded at a certain radius $\xi$
\begin{align}\label{eq:pinningpotential}
V_{\rm pin}^{\rm
V}=N\frac{V_0}{2}\left[\left(\frac{X}{\xi}\right)^2 +
\left(\frac{Z}{\xi}\right)^2\right]\theta\left(X^2
+Z^2-\xi^2\right)\!.
\end{align}

\subsection{Results}

\subsubsection{Zero temperature without extrinsic pinning}

When there are no external fields or pinning potentials, we can
read off the $T=0$ behavior directly from
Eqs.~(\ref{eq:langevinvortex1})~and~(\ref{eq:langevinvortex2}).
For instance, it is clear that under the influence of a current to
the right, the domain wall will move to the right also, and its
speed in the $x$-direction will be directly proportional to the
current, as can be seen in
\begin{align}
\dot{X} = \frac{1 + \frac{2 L_z}{\pi^2 \kappa} \alpha\beta}{1 +
\frac{2 L_z}{\pi^2 \kappa} \alpha^2} v_\s\; .
\end{align}
Note that if $\beta = \alpha$, the velocity of the domain wall
will be exactly equal to the velocity of the current. For the
transverse velocity $\dot{Z}$, we get
\begin{align}\label{eq:eomZ}
\dot{Z} = \frac{- s L_z(\beta - \alpha) v_\s}{\lambda \left(\pi +
\frac{2 L_z}{\pi \kappa} \alpha^2 \right)}\; .
\end{align}
The direction of motion in the $z$-direction depends on the
skyrmion number $s$, and also on the sign of $\beta - \alpha$, and
the magnitude is proportional to $v_\s$. Theoretically, if $\beta
= \alpha$, the vortex core would move in a straight line, i.e. the
center of the vortex would not get a transverse displacement. Note
that there is no intrinsic pinning in the case of a vortex domain
wall, not even if $\beta = 0$. For $\beta = \alpha = 0$,
$\langle\dot{X}\rangle = \frac{\lambda}{\kappa} v_\s$, showing
that the shape of the vortex domain wall has an influence on the
motion as well.

\subsubsection{Zero temperature with extrinsic pinning}

We will now investigate what happens if there is a pinning
potential of the form in Eq.~(\ref{eq:pinningpotential}). If the
potential is not bounded (i.e. if the step function is absent),
$T=0 $ and $H_{\rm ext}=0$, the solutions can be found
analytically, and they describe a circular motion ending at a
fixed point. As we can read from the formulas, these fixed points
will be at $X = \beta  v_\s\hbar\xi^2/(\lambda^2 V_0)$ and $Z = s
\pi v_\s \hbar  \xi^2/(\lambda L_z V_0)$. Note that only the
former depends on $\beta$. This agrees with our physical intuition
that $\beta$ tilts the potential landscape in the $X$-direction,
but that it has no effect in the $Z$-direction.

When the potential is bounded there is a critical current for
depinning the domain wall. We cannot find this current precisely
using analytic calculations, but we can however make a rough
estimation if we take the current for which the equilibrium
position falls outside the boundary as an indication. This will
naturally overestimate the critical current since the domain wall
precesses after the current is switched on, but it is a good
approximation for the upper boundary. We find

\begin{align}
v_{\rm s, crit} < \frac{\lambda^2 V_0 L_z}{\hbar \xi}
\sqrt{\frac{1}{L_z^2 \beta^2 + \lambda^2 \pi^2}}\;.\label{eq:beta}
\end{align}

\begin{figure}[h!]
\begin{center}
\includegraphics[width=8cm]{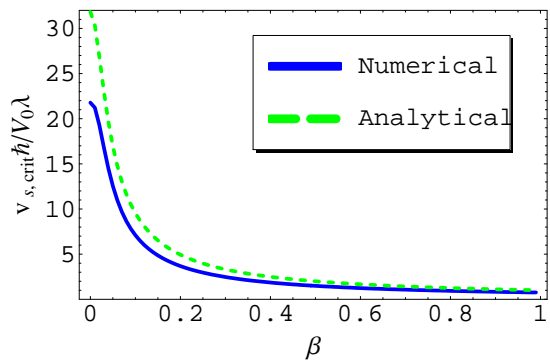}\\
\caption{(Color online) Critical current as a function of the
dissipative spin-transfer torque parameter
$\beta$.}\label{fig:beta}
\end{center}
\end{figure}

Numerical calculations can give us more precise results. For
instance, for a symmetric domain wall (i.e. $\kappa=\lambda$) with
$L_z = 100\lambda$, $\xi = \lambda$, and $\beta = 0$, we find
$v_{\rm s, crit} \simeq 21.78\lambda V_0/\hbar$, while the
analytical upper bound found with the formula above is $v_{\rm s,
crit} < 31.831 \lambda V_0/\hbar$. Numerical calculations also
give us the escape time at $v_{s,{\rm crit}}$, which is
approximately $0.074\hbar/V_0$ (for $\beta = 0$).
Eq.~(\ref{eq:beta}) suggests that the critical current is
proportional to $1/\beta$. Fig.~\ref{fig:beta} shows both
Eq.~(\ref{eq:beta}) and the numerical results for the
aforementioned values of $\kappa$, $L_z$ and $\xi$.

\subsubsection{Escape rate at finite temperature}

\begin{figure}[h!]
\begin{center}
\includegraphics[width=6.5cm]{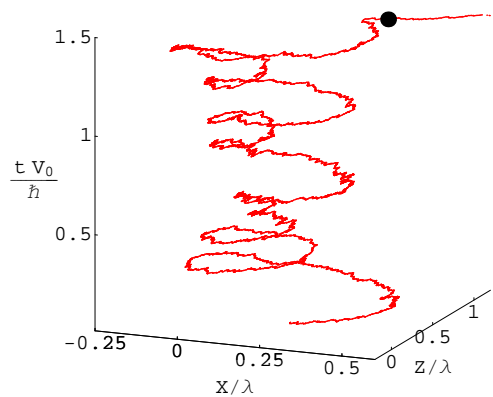}
\caption{A possible trajectory of a vortex domain wall in the
presence of the pinning potential given by
Eq.~(\ref{eq:pinningpotential}) with $\kB T L_z = N V\kappa$ and
$\kappa = \lambda = L_z/100$. The black dot denotes the
coordinates at which the vortex leaves the potential well, in this
example at time $t\simeq1.4\hbar/V_0$.} \label{fig:traj}
\end{center}
\end{figure}

We now include thermal fluctuations. To give some insight into the
motion of the domain wall in this case, we have plotted a possible
solution to the Langevin
Eqs.~(\ref{eq:langevinvortex1b})~and~(\ref{eq:langevinvortex2b})
in Fig.~\ref{fig:traj}. The current here is just under the
critical one, $\kB  T L_z = N V \kappa$, and we are looking at a
symmetric vortex with $L_z = 100 \lambda$.

\begin{figure}[!hbt]
\begin{center}
\includegraphics[width=8cm]{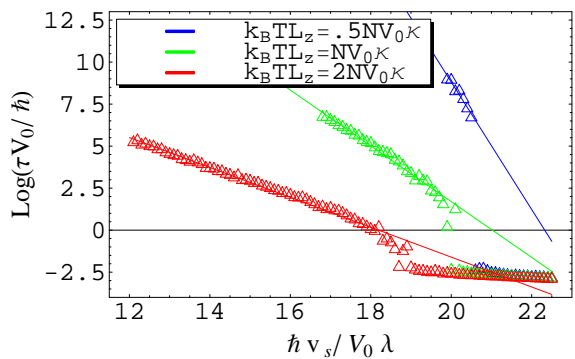}\\
\caption{Plot of the median escape time as a function of current,
for three different values of the temperature.}\label{fig:tau}
\end{center}
\end{figure}

We next determine the escape time (i.e. the time it takes for the
domain wall to move outside the boundary of the pinning potential)
as a function of the current, and its dependence on temperature.
As the average escape time is rather hard to determine and as the
escape times are distributed approximately exponentially, we have,
for practical purposes, chosen the median escape time as an
indicator. ${\rm Median}/\ln 2$ is then expected to be a good
measure for the average escape time. The results, with the
logarithm of the median escape time plotted against the current,
are shown in Fig.~\ref{fig:tau}.

Note that there is a clear distinction between the thermal regime
and the slide regime. The behavior in the thermal regime can be
fitted very well by an equation of the form $\exp[(a - b
v_\s/v_{\rm s, crit}) N V \kappa /\kB  T L_z]$, with $a$ and $b$
two numerical factors. Fitting shows $a$ to be of the order of
$35$, and $b$ around $1.8$. Hence we find that the logarithm of
the escape time is proportional to current.

\subsection{Comparison with other work}

Vortex domain walls are of great interest to experimentalists,
because they are large enough to be visible with, for instance,
scanning electron microscopy or magnetic force microscopy, and
because their dynamics take place on observable time scales.
Because many experiments are done with vortex domain walls, many
theoretical models have been developed to describe them. He, Li
and Zhang\cite{he2006} developed a 2D model for the vortex domain
wall, while Shibata {\it et al.} \cite{shibata2006} described the
current-induced motion of a circular magnetic vortex. The biggest
difference between these two models and ours is that in their
case, the vortex (domain wall) is not straight, but the equations
of motion they find are remarkably similar to ours in the
isotropic, symmetric case. The former group found the motion of
the domain wall towards the edge of the nanostrip that our model
also yields, and the latter paper predicted precession of the
vortex domain wall in a potential. The model of Kr\"{u}ger {\it et
al.}\cite{kruger2007} also predicts precession, of an elliptical
shape, and our zero-temperature results are in agreement with
theirs.

Current-driven vortex domain wall motion has indeed been observed
by several experimental groups.\cite{klaui2005, yamaguchi2004,
thomas2006} Furthermore, Heyne {\it et al.} \cite{heyne2008}
found, as expected, that the sign of the displacement in the
$Z$-direction is determined by the skyrmion number and by the sign
of $(\alpha - \beta)$. In the presence of pinning, precession of a
magnetic vortex has indeed been found experimentally as
well.\cite{bolte2008}

No detailed experimental results on current-driven magnetic
vortices at nonzero temperature have been reported.

\section{Conclusions}

We have presented a model for the driven motion of a domain-wall
line at nonzero temperature, and analyzed this model within
several approximations.

First, we considered a general domain-wall line, which is
described as a string in a tilted washboard potential. We computed
the average drift velocity as a function of an applied current for
currents lower than the critical current.

In the limit of rigid domain walls, we were able to find
analytical expressions for drift velocities in the presence of
thermal fluctuations. For the field-driven case, the well-known
Walker-breakdown behavior smoothes for increasing temperature. In
the current-driven case, the drift velocity depends heavily on the
ratio of the dissipative spin-transfer torque parameter $\beta$
and the Gilbert damping $\alpha$. Here, the curves also smoothen
with increasing temperature. As a result, we found no critical
current for $\beta=0$ at nonzero temperature. We also considered
extrinsic pinning due to, for example, a notch in the sample. We
found that the escape time is proportional to the exponent of the
applied magnetic field and the dissipative spin-transfer torque.
Comparison with experiment enabled us to estimate
$\beta\sim10^{-2}$ for that experiment\cite{ravelosona2005}.
Finally, we discussed the effect of a disorder potential on the
dynamics of the rigid domain wall.

For vortex domain walls, we computed domain-wall velocities at
zero temperature. In the presence of an extrinsic pinning
potential, we found an analytic upper bound for the critical
current. Numerical computation revealed critical currents just
under this upper bound. At finite temperature, numerical
simulations provided us with depinning times as a function of the
applied current. We found two distinct regimes: one where thermal
fluctuations dominate, and one where the current dominates. In the
thermal regime, we found that the depinning time goes as
$\log{\tau}\propto v_\s/T$.

The models and results presented in this paper provide a simple
framework for describing domain walls at nonzero temperature.
Moreover, they are easily adapted to situations not discussed in
this paper, such as different geometries. We hope that our results
are confirmed with more experimental and numerical results in the
near future.

\begin{acknowledgments}
This work was supported by the Netherlands Organization for
Scientific Research (NWO), by the European Research Council (ERC)
under the Seventh Framework Program (FP7), and by the National
Science Foundation under Grant No. NSF PHY05-51164.
\end{acknowledgments}


\begin{thebibliography}{99}

\bibitem{berger1984} L. Berger, J. Appl. Phys. {\bf 55}, 1954
(1984).

\bibitem{berger1985} P.P. Freitas and L. Berger, J. Appl.
Phys. {\bf 57}, 1266 (1985).

\bibitem{slonczewski1996} J.C. Slonczewski, J. Magn. Magn. Mater.
{\bf 159}, L1 (1996).

\bibitem{berger1996} L. Berger, Phys. Rev. B. {\bf 54}, 9353
(1996).

\bibitem{zhang2004} S. Zhang and Z. Li, Phys. Rev. Lett. {\bf 93},
127204 (2004).

\bibitem{barnes2005} S.E. Barnes and S. Maekawa, Phys. Rev. Lett.
{\bf 95}, 107204 (2005).

\bibitem{heyne2008} L. Heyne {\it et al.},
Phys. Rev. Lett. {\bf 100}, 066603 (2008).

\bibitem{duine2007a} R.A. Duine, A.D. N\'{u}\~{n}ez, J. Sinova and
A. H. MacDonald, Phys. Rev. B {\bf 75}, 214420 (2007).

\bibitem{tserkovnyak2006} T. Tserkovnyak, H.J. Skadsem, A. Brataas
and G.E.W. Bauer, Phys. Rev. B {\bf 74}, 144405 (2006).

\bibitem{kohno2006} H. Kohno, G. Tatara and J. Shibata, J. Phys.
Soc. Jpn. {\bf 75}, 113706 (2006).

\bibitem{piechon2007} F. Pi\'echon and A. Thiaville, Phys. Rev. B
{\bf 75}, 174414 (2007).

\bibitem{petit2007}  D. Petit, A.-V. Jausovec, D. Read, and
R.P. Cowburn, J. Appl. Phys. {\bf 103}, 114307 (2008).

\bibitem{ravelosona2005} D. Ravelosona, D. Lacour, J. A. Katine,
B. D. Terris and C. Chappert, Phys. Rev. Lett. {\bf 95}, 117203
(2005).

\bibitem{bruno1999} P. Bruno, Phys. Rev. Lett. {\bf 83}, 2425
(1999).

\bibitem{shibata2006} J. Shibata, Y. Nakatani, G. Tatara,
H. Kohno and Y. Otani, Phys. Rev. B {\bf 73}, 020403 (R) (2006).

\bibitem{he2006} J.He, Z.Li and S. Zhang,
Phys. Rev. B. {\bf 73}, 184408 (2006).

\bibitem{kruger2007} B. Kr\"uger, A. Drews, M. Bolte,
U. Merkt, D. Pfannkuche and G. Meier, Phys. Rev. B {\bf 76},
224426 (2007).

\bibitem{hayashi2006} M. Hayashi, L. Thomas,
C. Rettner, R. Moriya, X. Jiang, and S.S.P. Parkin, Phys. Rev.
Lett. {\bf 97}, 207205 (2006).

\bibitem{bolte2008} M. Bolte {\it et al.},
Phys. Rev. Lett. {\bf 100}, 176601 (2008).

\bibitem{klaui2005} M. Kl\"aui, P.-O. Jubert, R. Allenspach,
A. Bischof, J. A. C. Bland, G. Faini, U. R\"udiger, C. A. F. Vaz,
L. Vila and C. Vouille, Phys. Rev. Lett. {\bf 95}, 026601 (2005).

\bibitem{duine2008} R.A. Duine and C. Morais Smith,
Phys. Rev. B {\bf 77}, 094434 (2008).

\bibitem{duine2007b} R.A. Duine, A.S. N\'{u}\~{n}ez and A.H.
MacDonald, Phys. Rev. Lett. {\bf 98}, 056605 (2007).

\bibitem{lecomte2009} V. Lecomte, S.E. Barnes, J.-P. Eckmann and
T. Giamarchi, cond-mat.stat-mech/0903.0175 (2009).

\bibitem{tatara2005} G. Tatara, N. Vernier and J. Ferr\'e, Appl.
Phys. Lett. {\bf 86}, 252509 (2005).

\bibitem{martinez2007} E. Martinez, L. Lopez-Diaz, O. Alejos,
L. Torres and C. Tristan, Phys. Rev. Lett. {\bf 98}, 267202
(2007).

\bibitem{tatara2004} G. Tatara and H. Kohno, Phys. Rev. Lett. {\bf
92}, 086601 (2004); {\bf 96}, 189702 (2006).

\bibitem{riskenbook} H. Risken, {\it The Fokker-Planck Equation}
(Springer-Verlag, Berlin, 1984).

\bibitem{buttiker1981} M. B\"uttiker and R. Landauer, Phys. Rev. A
{\bf 23}, 1397 (1981).

\bibitem{yamanouchi2007} M. Yamanouchi, J. Ieda, F. Matsukura,
S. E. Barnes, S. Maekawa, H. Ohno, Science {\bf 317}, 1726 (2007).

\bibitem{moore2008}  T.A. Moore, I.M. Miron, G. Gaudin,
G. Serret, S. Auffret, B. Rodmacq, A. Schuhl, S. Pizzini, J. Vogel
and M. Bonfim, cond-mat.other/0812.1515 (2008).

\bibitem{walker1974} N.L. Schryer and L.R. Walker,
J. Appl. Phys. {\bf 45}, 5406 (1974).

\bibitem{beach2005} G.S.D. Beach, C. Nistor, C. Knutson,
M. Tsoi and J.L. Erskine, Nature Mat. {\bf 4}, 741 (2005).

\bibitem{ledoussal1995} P. Le Doussal and V.M. Vinokur,
Physica C {\bf 254}, 63 (1995).

\bibitem{yamaguchi2004} A. Yamaguchi, T. Ono, S. Nasu,
K. Miyake, K. Mibu and T. Shinjo, Phys. Rev. Lett. {\bf 92},
077205 (2004).

\bibitem{thomas2006} L. Thomas, M. Hayashi, X. Jiang, R. Moriya,
C. Rettner and S.S.P. Parkin, Nature {\bf 443}, 197 (2006).

\end{thebibliography}
\end{document}